\documentclass[%
 reprint,
 amsmath,amssymb,
 aps,
prstab,
floatfix,
]{revtex4-1}

\usepackage{graphicx}
\usepackage{dcolumn}
\usepackage{bm}
\usepackage{lipsum}
\usepackage{verbatim}
\usepackage{color}

\newcommand{\DE}{\Delta_\textrm{E}}

\newcommand{\Ud}{U_\textrm{d}}
\newcommand{\lamo}{\lambda_\textrm{0}}

\newcommand{\Eo}{E_\textrm{0}}
\newcommand{\Np}{N_\textrm{p}}

\begin{document}


\title{Reconfigurable Photonic Circuit for Controlled Power Delivery to Laser-Driven Accelerators on a Chip}

\author{Tyler W. Hughes}
\affiliation{Department of Applied Physics, Stanford University, Stanford, CA, 94305.}
\author{R. Joel England}
\affiliation{SLAC National Accelerator Laboratory, Menlo Park, CA, 94025.}
\author{Shanhui Fan}%
 \email{shanhui@stanford.edu}
\affiliation{Ginzton Laboratory, Stanford University, Stanford, CA, 94305.}


\date{\today}

\begin{abstract}
Dielectric laser acceleration (DLA) represents a promising approach to building miniature particle accelerators on a chip. However, similar to conventional RF accelerators, an automatic and reconfigurable control mechanism is needed to scale DLA technology towards high energy gains and practical applications.  We present a system providing control of the laser coupling to DLA using integrated optics and introduce a novel component for power distribution using a reconfigurable mesh of Mach-Zehnder interferometers.  We show how such a mesh may be sequentially and efficiently tuned to optimize power distribution in the circuit and find that this strategy has favorable scaling properties with respect to size of the mesh.
\end{abstract}

\maketitle

\section{\label{sec:intro}Introduction}

Dielectric laser acceleration (DLA) is a promising candidate for the next generation of particle accelerators, which makes use of advances in precision nanofabrication, high power lasers, and integrated optics \cite{peralta_demonstration_2013,breuer_laser-based_2013,breuer_dielectric_2014,leedle_dielectric_2015,leedle_laser_2015,wootton_demonstration_2016}.  In DLA, infrared lasers are used to power micron-scale lithographically fabricated dielectric structures, which are designed to create an electromagnetic field pattern that provides extended energy gain to charged particles that traverse the device \cite{plettner_proposed_2006, hughes_method_2017}.  DLA offers orders of magnitude increases in the achievable energy gain per length, or `acceleration gradient', mainly because of the high damage thresholds of dielectric materials when compared to the metal used in conventional Radio Frequency (RF) accelerators \cite{soong2012laser}.  The ability to construct compact, inexpensive, and powerful particle accelerators would have numerous applications \cite{england_dielectric_2014,wootton_dielectric_2016,wootton_towards_2017}, especially if the technology may be implemented on a chip using an integrated photonic circuit \cite{hughes2018chip}.

A major challenge of DLA is increasing the length of interaction between the driving laser and the electron beam, which is limited by both the electron beam dynamics and the laser delivery system.  Without an acceleration length at or above the millimeter scale, energy gains from DLA will remain below $1$ MeV, which is not enough for most practical applications.  A promising approach to scaling the interaction length involves using integrated optics platforms, built with precise nanofabrication, to provide controlled laser power delivery to the DLA.  This scheme would eliminate many free-space optical components, which are bulky, expensive, and sensitive to alignment.  Furthermore, as de-phasing between the electron beam and the driving laser is a large barrier to long interaction lengths, an integrated laser delivery and control mechanism would mitigate this issue by allowing automatic, reconfigurable control over the accelerator structure, similar to what is commonplace in modern RF accelerators.

A system for accomplishing integrated optical power delivery was recently proposed in Ref. \cite{hughes2018chip}, in which the laser beam is first coupled into a single dielectric waveguide and then split into several waveguides that spread to power the length of the accelerator structure.  Here, waveguide bends are designed to implement an on-chip `pulse-front tilt', which delays the laser pulses in each waveguide to arrive at the accelerator structure at the same time as the moving electron beam \cite{cesar_optical_2018}.  However, this design has two serious limitations.  First, the design relies on a single input facet to couple the laser beam to a single waveguide before it is split.  The optical power concentration in the input facet thus creates a bottleneck leading to optical damage and nonlinearities, which limits the maximum input power and, in turn, energy gain of the accelerator.  Secondly, while the phase of each output pulse is tuned using integrated optical phase shifters, the power distribution of the output waveguides is fixed by the fabrication of the splits and bends in the structure and may not be easily tuned later if there are errors.

Therefore, an ideal laser coupling scheme for DLA would involve a `one-to-many' coupling scheme that directly couples a single laser pulse into several waveguides.  This would greatly ease the input power limitations present in the power-splitting proposal of Ref. \cite{hughes2018chip} by eliminating the optical damage and nonlinearity bottleneck of the input coupler.  While `one-to-many' coupling techniques exist in the form of grating coupler arrays \cite{van2011optical} or combined grating coupler and splitter devices \cite{spuesens2016grating}, a tunable $\textit{power}$ distribution system would also be necessary to mitigate the variations in input coupling efficiency, fabrication errors, and drift experienced during the experiment.  In conjunction with optical phase control and long-range beam-focusing \cite{niedermayer_alternating_2018}, a reconfigurable power distribution mechanism would be an essential component for scaling DLA from proof-of principle length scales of 100's of $\mu$m to the scale of modern accelerators.

In this work, we propose an integrated power distribution system and corresponding protocol for accomplishing automated, on-chip phase and power control for DLA and other applications. For this, we utilize a mesh of integrated Mach-Zehnder interferometers (MZIs), a device that allows for reconfigurable unitary operations on-chip \cite{miller_self-configuring_2013,miller_perfect_2015}.  MZI meshes are becoming a fundamental component in integrated, reconfigurable optics for mode sorting \cite{miller_sorting_2015,annoni_unscrambling_2017,miller_setting_2017,miller_self-configuring_2018}, quantum information processing \cite{harris_quantum_2017,metcalf_multiphoton_2013,aspuru-guzik_photonic_2012,obrien_photonic_2009}, and optical machine learning \cite{shen_deep_2017, hughes2018training}.  Here, we explore the novel application of these reconfigurable optics systems to high power pulse delivery and control for accelerators on a chip, which brings its own interesting set of constraints and challenges.  In conjunction with Ref. \cite{hughes2018chip}, this study gives a roadmap for transitioning the control of DLA systems away from hand-tuned, free-space optical setups to precise, automatically configured integrated optical components.  As such, our work points to the opportunity for the use of on-chip reconfigurable optics to dramatically scale up the acceleration length and energy gains of DLA, which is of crucial importance in realizing the exciting applications of these miniature accelerators.

The paper is organized as follows: In Section \ref{sec:system}, we give a systems-level overview of the proposed laser coupling system and we introduce the MZI mesh as a power distribution component for DLA.  In section \ref{sec:algo} we provide a sequential protocol for optimizing the mesh for delivery to the accelerator, which involves layer-wise tuning of individual MZIs using photodetector measurements.  In section \ref{sec:demo}, we perform numerical simulations of reconfiguring such a mesh to demonstrate our method. Then, in Section \ref{sec:DLA_analysis}, we calculate the improvement of the DLA performance with our approach using experimentally measured parameters.  In Section \ref{sec:discussion}, we discuss the scalability of our approach and other considerations before concluding in Section \ref{sec:conclusion}.

\section{\label{sec:system}Reconfigurable control system}

\begin{figure}
\includegraphics[width=\columnwidth]{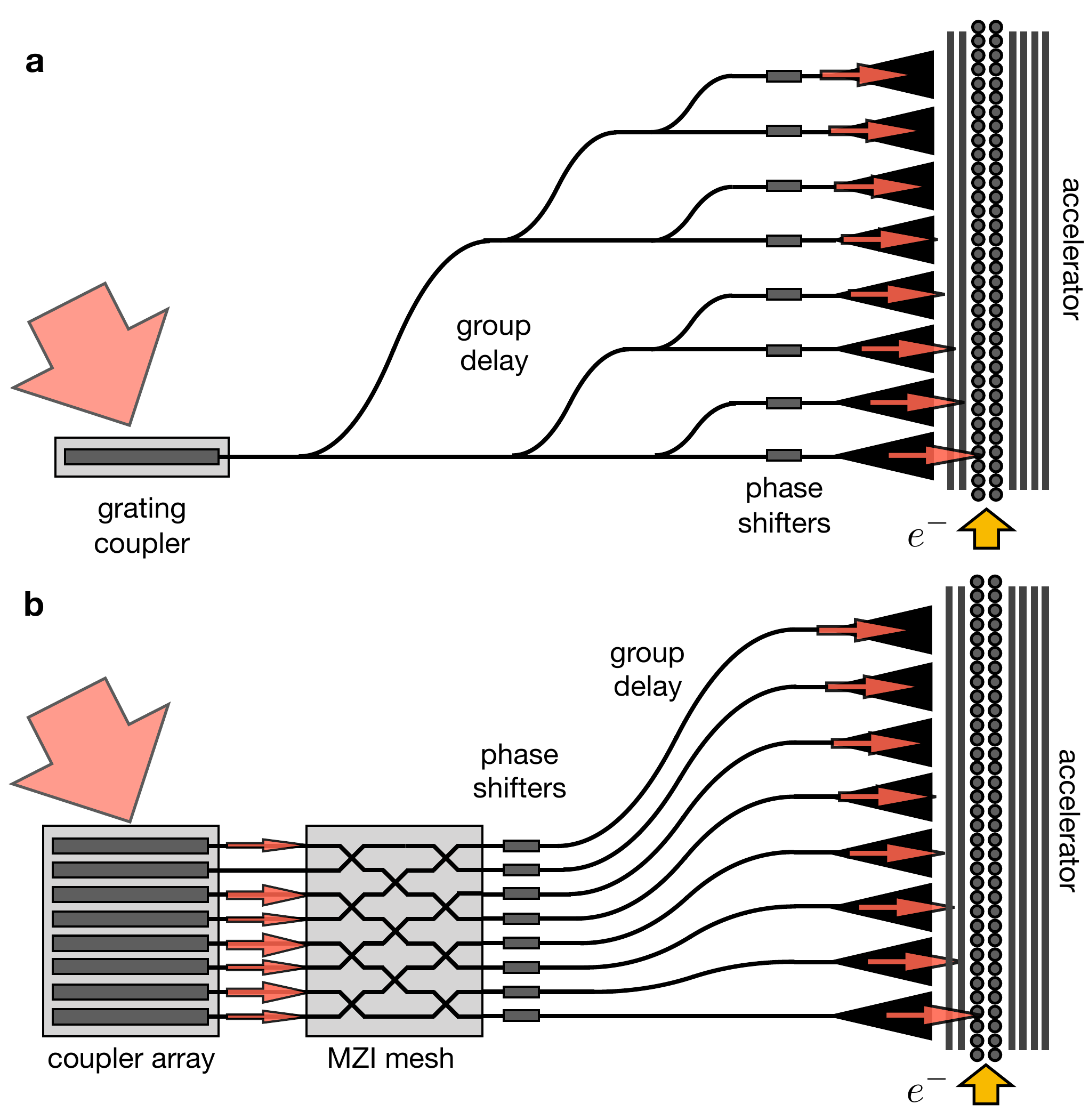}
\caption{\label{fig:setup} \textbf{Schematic of the proposed DLA laser delivery control system.}  Comparison of setup from Ref \cite{hughes2018chip} (\textbf{a}) and this work (\textbf{b}).  \textbf{a}, In the splitting structure of Ref. \cite{hughes2018chip}, an input pulse (large red arrow) is focused to a single grating coupler on the chip surface.  The pulse is split a series of times and bends in the waveguide create group velocity delay.  The phase of the pulses are corrected before injection into the accelerator.  \textbf{b}, In the schematic proposed in this work, the pulse is directly coupled into several waveguides through a grating coupler array. To mitigate the large variance in coupled powers, the pulses next enter a mesh of MZIs, which is sequentially adjusted, using the protocol from this paper, to provide uniform powers in each waveguide.  As before, phase control and group delay are performed using integrated phase shifters and lithographically-defined bends before couping to the accelerator channel.  The second scheme eliminates the damage and nonlinearity bottleneck present in $\textbf{a}$ directly after the input coupler.}
\end{figure}

Here we will give an overview of the proposed reconfigurable laser coupling scheme for DLA.  A schematic of our system is shown in Fig. \ref{fig:setup}.  The system consists of sequential stages for input coupling, power distribution, phase control, group delay control, and electron acceleration. 

In our design, the driving laser pulse is first focused onto an input element that directly splits the optical power into several waveguides.  Compared with schemes where the laser is first coupled to a single waveguide, as in Ref \cite{hughes2018chip}, this coupling strategy can greatly improve the power that can be safely supplied by the driving laser. This element could take the form of a grating coupler array or combined grating coupler and power splitter geometry \cite{spuesens_grating_2016}.  Adjoint-based optimization techniques \cite{sapra2019inverse} may be employed to design novel input coupling components with improved coupling efficiency, less variation between powers, or significantly more output waveguides, although this is outside the focus of this work.

\begin{figure}
\includegraphics[width=1\columnwidth]{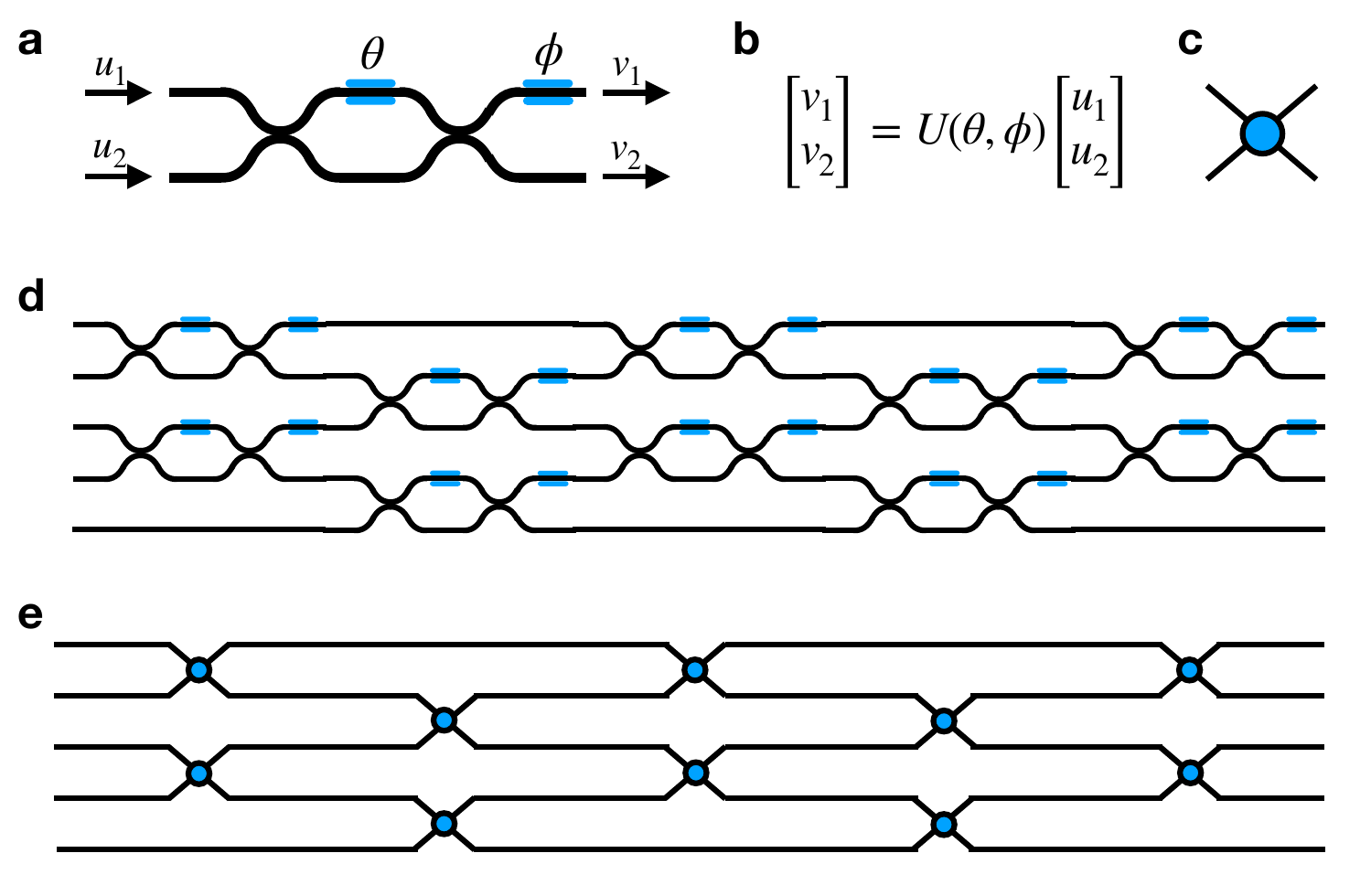}
\caption{\label{fig:mesh} \textbf{Diagram of MZI mesh for power distribution.} \textbf{a}, Diagram of a single MZI consisting of two input ports and two output ports.  Where the two arms come together, a 50-50 beam-splitter operation is performed.  The blue regions indicate tunable optical phase shifters with added phases marked as $\theta$ and $\phi$.  \textbf{b}, The MZI represents a tunable unitary transformation on its inputs $[u_1, u_2]^T$ to give outputs $[v_1, v_2]^T$.  \textbf{c}, Simplified diagram representing an MZI in the following figures.  One may think of this element as a tunable power switch. \textbf{d}, Individual MZIs are combined into meshes, which may implement tunable unitary operations on several inputs.  Shown is a 'Clements' Mesh geometry, which was used in this work. \textbf{e}, Schematic of the Clements mesh in \textbf{d} using the simple MZI diagram in \textbf{c}.}
\end{figure}

After coupling, due to fabrication and alignment errors, there will be much variation in the power distribution of each waveguide. To ensure that an equal amount of power is supplied to the accelerator, we introduce a power distribution component that is comprised of a mesh of MZIs.  As diagrammed in Fig. \ref{fig:mesh} \textbf{a}-\textbf{c}, each individual MZI is comprised of two beam-splitters and two optical phase shifters, $\theta$ and $\phi$, which can be electrically adjusted to perform the following unitary operation on its two inputs \cite{reck_experimental_1994,clements_optimal_2016,shen_deep_2017,pai2018matrix}
\begin{align}
\begin{split}
    \begin{bmatrix}
      v_1 \\ v_2
    \end{bmatrix}
    =&~U(\theta, \phi) 
    \begin{bmatrix}
      u_1 \\ u_2
    \end{bmatrix}
    \\
    =&~e^{i\frac{\theta}{2}} e^{i\frac{\phi}{2}}
    \begin{bmatrix}
      e^{i\frac{\phi}{2}}\cos{\frac{\theta}{2}} &
      e^{i\frac{\phi}{2}}\sin{\frac{\theta}{2}} \\
      - e^{-i\frac{\phi}{2}}\sin{\frac{\theta}{2}} & 
      e^{-i\frac{\phi}{2}}\cos{\frac{\theta}{2}}
    \end{bmatrix}
    \begin{bmatrix}
      u_1 \\ u_2
    \end{bmatrix}.
\end{split}
\label{eq:U}
\end{align}

As shown in Figs. \ref{fig:mesh} \textbf{d},\textbf{e}, several MZIs may be combined in a mesh, which is capable of performing unitary operations over an arbitrary large number of inputs.  There are several possible configurations of the MZI mesh, each with their own benefits and drawbacks.  In this application, the mesh must be compact enough to fit on the chip and also have a bandwidth large enough to handle sub-picosecond pulses.  Whereas many meshes are capable of performing arbitrary unitary operations, for this power delivery problem it is only necessary to sort a single random input into a uniform output.  With these considerations in mind, the `Clements' mesh geometry \cite{clements_optimal_2016}, in which MZIs are configured in a rectangular mesh, is best suited for this application.   As diagrammed in Fig. \ref{fig:mesh} \textbf{d},\textbf{e}, the Clements mesh requires fewer layers than other designs \cite{reck_experimental_1994} and also is more robust to optical losses because of its symmetric layout.  Furthermore, it may be implemented in a `shallow` mesh with fewer layers than input ports.  This is especially useful for DLA power distribution problem as we will show that only a few layers are required for adequate power equalization.  

In the following Section we describe a sequential protocol for optimizing the Clements mesh for power distribution. As our protocol uses local information about the power within the network, we require the inclusion of integrated optical photodetectors within the MZI mesh, such as those used in \cite{annoni_unscrambling_2017}.  Alternatively, an imaging system, in conjunction with scattering elements, may be used to gather information about the power distribution from above the chip.

Once the power in each waveguide is equalized, we may use integrated optical phase shifters to correct the phase of each laser pulse such that it is synchronous with the electron beam. While the phase shifters may be adjusted to maximize energy gain, they may also be used to incorporate other functionality, such as beam focusing, total beam transmission, deflection, or diagnostics.  Beam energy measurements, performed periodically along the accelerator, may be used as a signal to automatically configure the optical phase shifters.

Once the power and phase are sufficiently controlled, we must delay the laser pulse in each waveguide so that it arrives at the accelerator gap the same time as the moving electron beam. To do this, we introduce lithographically-defined bends in the waveguides to provide a delay that is matched to the electron arrival, producing of the integrated optics analogue of a pulse-front tilt \cite{cesar_optical_2018}. The mathematical details of the bend design are described in the supplementary material of Ref. \cite{hughes2018chip}.

The final stage involves coupling the waveguide mode into the acceleration channel.  Ideally, an inverse taper may be placed at the end of each waveguide to spread the mode area to match the geometry of the electron beam.  Then, an accelerator structure may be placed adjacent to the end of the waveguides.  Alternatively, the accelerator structures and tapers may be part of the same system and may be designed following ideas from proposed buried grating structures \cite{chang_silicon_2014}, or using inverse design techniques, such as what was shown Ref. \cite{hughes_method_2017} for free-space coupling.  Dielectric mirrors may be used to design the resonance of the acceleration cavity and provide back reflection if a single-sided drive is used (as pictured in Fig. \ref{fig:setup}).  While Ref. \cite{hughes2018chip} argued that a moderately resonant acceleration cavity with quality factor of several hundred would be necessary to lower the input power to mitigate the input facet bottleneck introduced by that design.  In this proposal, because that bottleneck is eliminated, one might not require such a resonant structure.

\section{\label{sec:algo}Power Distribution Protocol}

With the control system defined, here we describe a power distribution protocol that may be implemented on a Clements mesh to optimally equalize the power coming from a random input source.  The goal of this power equalization stage is to find the settings of each of the integrated phase shifters such that, given an input to the mesh, there is an equal power in each output port.  A naive implementation of this may involve performing a global optimization over each of the phase shifters.  However, when the number of degrees of freedom increases (for example, for a very large accelerator), this becomes unfeasible as the dimension of the search space scales linearly with the number of MZIs. Fortunately, the protocol presented in this work allows each MZI to be tuned individually and in sequence, layer by layer, from input to output.

\begin{figure}[htp]
\includegraphics[width=0.7\columnwidth]{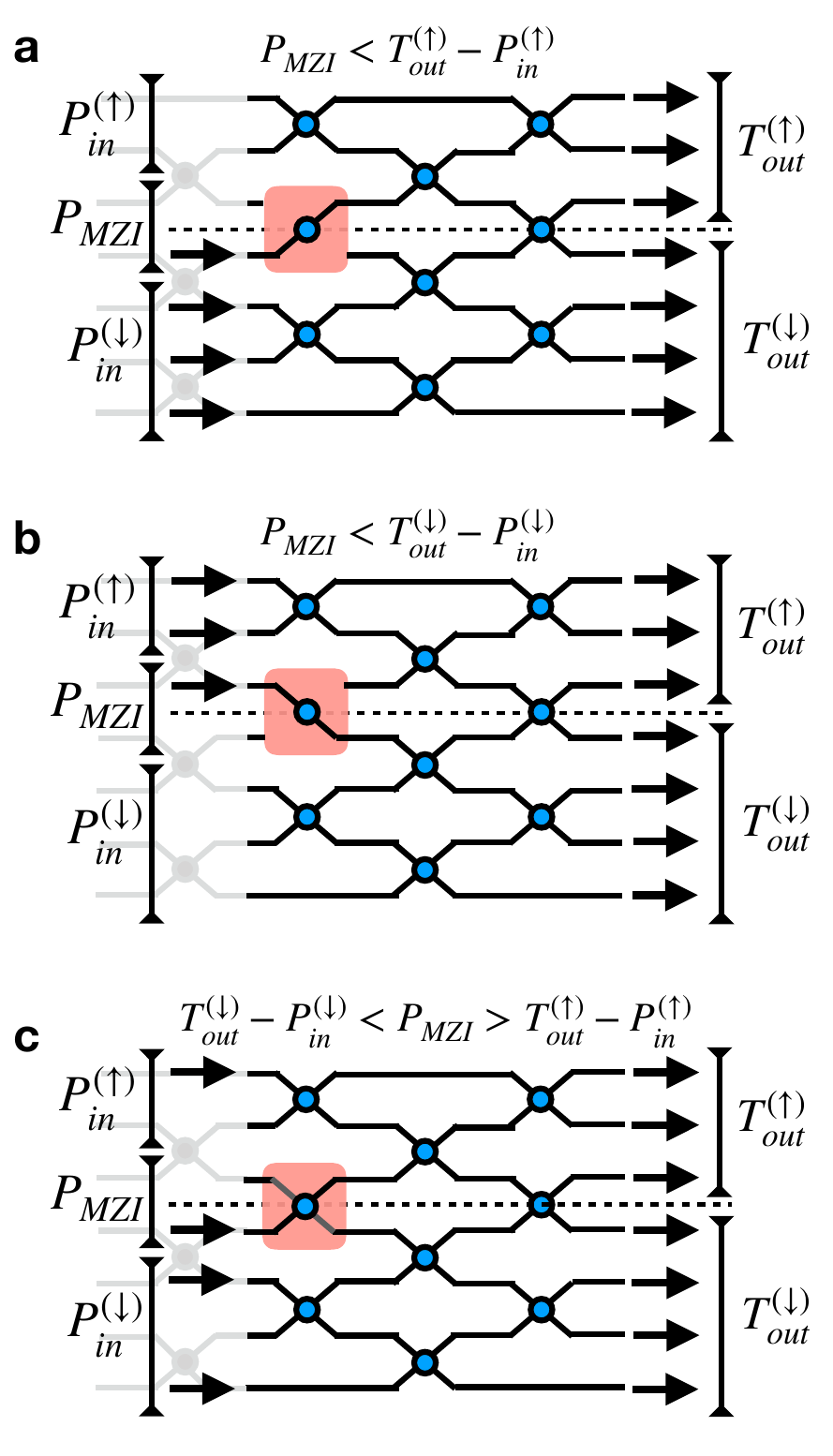}
\caption{\label{fig:algo} \textbf{Sequential algorithm for tuning single MZI.} The red square outlines the MZI being tuned.  Black arrows on the left indicate input power to the current layer, black arrows on the right indicate desired output powers (uniform in this case).  The dotted line separates the mesh into powers above and below the MZI.  \textbf{a} When there is more power needed in the ports above the MZI than are supplied in the ports leading into the MZI and above, the MZI should output all of its power to its top output port. \textbf{b} When there is more power needed in the ports below the MZI than are supplied in the ports leading into the MZI and below, the MZI should send all of its power to the bottom output port.  \textbf{c} In the intermediate case, the MZI should output just enough power to its top and bottom output ports to match the target power requirements.}
\end{figure}

Our protocol is diagrammed in Fig. \ref{fig:algo}, in which we show how to tune a single MZI (red) in a given layer of the network based purely on information about the powers coming into that layer.  For generality, we assume that we wish to tune this mesh to achieve an arbitrary target output power distribution $T_\textrm{out}^{(i)}$ for each of the $N$ ports $i~\in~\{1~..~N\}$.  Here in Fig. 3 the top port is at index $1$ and the bottom port is at index $N$.  We also assume that we have knowledge of the power at each of the ports coming into this layer, labelled $P_\textrm{in}^{(i)}$.

The essential idea of the protocol is to tune each MZI to locally direct power to either its top output port or its bottom output port depending on where power is deficient in the input to this layer and where it is needed in the final output layer. To visually represent this idea, in Fig. \ref{fig:algo}, we show a horizontal line bisecting the mesh through the MZI in question.  Assuming that the MZI is located vertically within the mesh with its top input  port at index $j$, we may define the sum of power input to this specific MZI as
\begin{equation}
    P_\textrm{MZI} \equiv P_\textrm{in}^{(j)} + P_\textrm{in}^{(j+1)}.
\end{equation}
The sum of the powers input to this layer both above and below this MZI, respectively, are defined as
\begin{align}
    P_\textrm{in}^{(\uparrow)} &\equiv \sum_{i=1}^{j-1} P_\textrm{in}^{(i)}\\
    P_\textrm{in}^{(\downarrow)} &\equiv \sum_{i=j+2}^N P_\textrm{in}^{(i)}.
\end{align}
Finally, the target powers above and below this MZI, respectively, are defined as
\begin{align}
    T_\textrm{out}^{(\uparrow)} &\equiv \sum_{i=1}^j T_\textrm{out}^{(i)} \\
    T_\textrm{out}^{(\downarrow)} &\equiv \sum_{i=j+1}^N T_\textrm{out}^{(i)}.
\end{align}
Now, using these values, we give a prescription for directing the power out of the MZI to optimally match the target.  We notice that there are three distinct cases to consider.  These are each diagrammed separately in the subplots of Fig. \ref{fig:algo}.  

Case 1 (Fig. \ref{fig:algo}\textbf{a}):  the sum of power supplied to the MZI and the ports \textit{above} it is less than the sum of power needed at the final output \textit{above} the MZI.  This is also written as
\begin{equation}
    P_\textrm{MZI} < T_\textrm{out}^{(\uparrow)} -  P_\textrm{in}^{(\uparrow)}.
\end{equation}
In this case, we require that the MZI direct all of its power to the \textit{top} output port, as shown in the red box of Fig. \ref{fig:algo}\textbf{a}.

Case 2 (Fig. \ref{fig:algo}\textbf{b}):  the sum of power supplied to the MZI and the ports \textit{below} it is less than the sum of power needed at the final output \textit{below} the MZI.  This is also written as
\begin{equation}
    P_\textrm{MZI} < T_\textrm{out}^{(\downarrow)} - P_\textrm{in}^{(\downarrow)}.
\end{equation}
In this case, we require that the MZI directs all of its power to the \textit{bottom} output port, as shown in the red box of Fig. \ref{fig:algo} \textbf{b}.

Case 3 (Fig. \ref{fig:algo}\textbf{c}):  when neither of the two cases above are satisfied, i.e.:
\begin{align}
\begin{split}
    P_\textrm{MZI} &\geq T_\textrm{out}^{(\uparrow)} -  P_\textrm{in}^{(\uparrow)} ~~~~~\textrm{and} \\
    P_\textrm{MZI} &\geq T_\textrm{out}^{(\downarrow)} - P_\textrm{in}^{(\downarrow)},
\end{split}
\end{align}
we require the MZI only supply $T_\textrm{out}^{(\uparrow)} - P_\textrm{in}^{(\uparrow)}$ of its power to the top port.  The leftover power may be transmitted to the down port, which, by power conservation, will be equal to $T_\textrm{out}^{(\downarrow)} - P_\textrm{in}^{(\downarrow)}$ since $\sum_{i=1}^{N}P_\textrm{in}^{(i)} = \sum_{i=1}^{N}T_\textrm{out}^{(i)}$.  This is demonstrated in Fig \ref{fig:algo}\textbf{c} where the MZI performs a partial splitting of power.

With this protocol, one may thus optimize each MZI sequentially through the mesh.  To do this optimally, the MZIs must be tuned layer-by-layer from input to output.  Within each layer, a set of integrated photodetectors must be used to measure $P_\textrm{in}^{(i)}$ for all ports $i$.  Then, the individual MZIs in this layer may be tuned in parallel or in any order desired.

\section{\label{sec:demo}Numerical Demonstration}

\begin{figure}[htp]
\includegraphics[width=1\columnwidth]{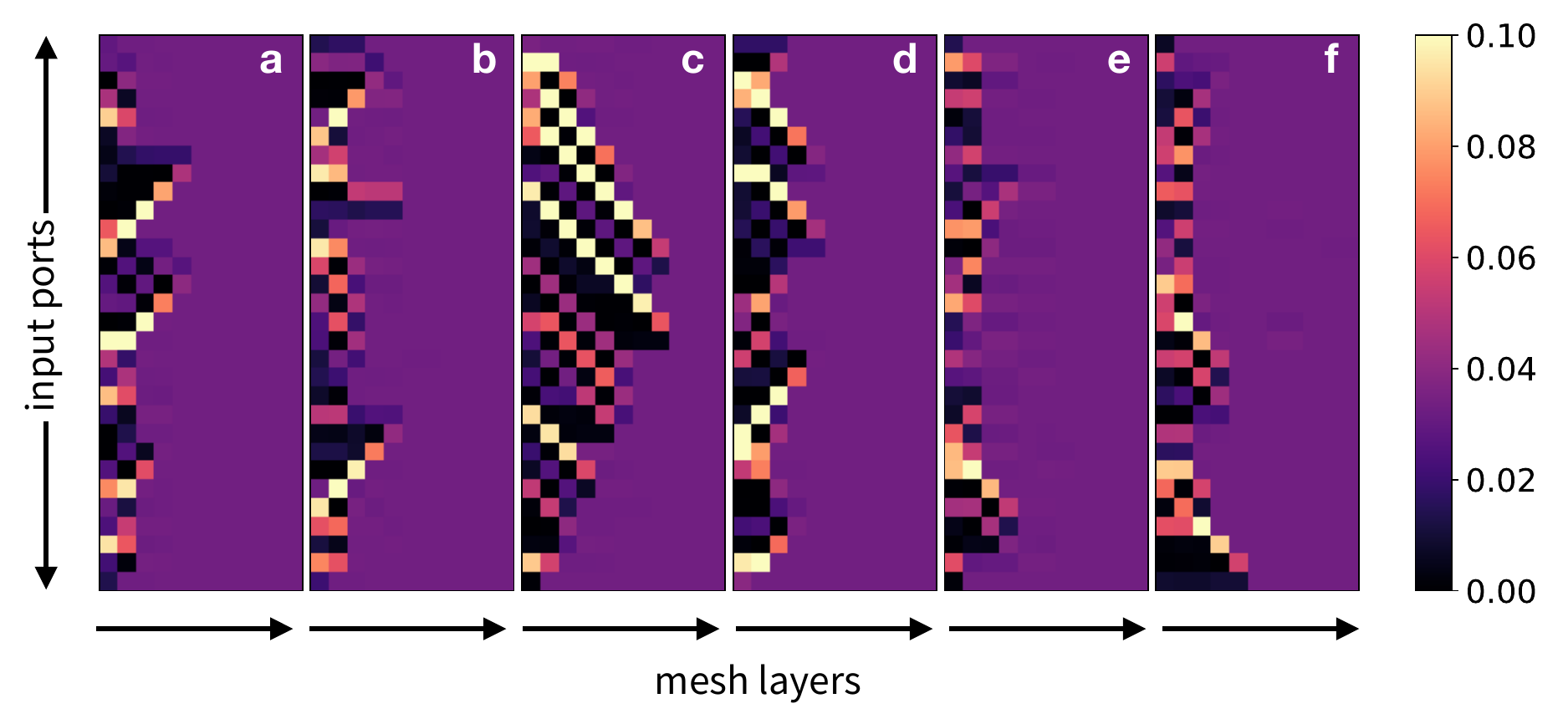}
\caption{\label{fig:equalization} \textbf{Demonstration of mesh optimization for equal power distribution.} \textbf{a}-\textbf{f} The powers in each port after optimizing the mesh for uniform output power given several random input powers.  The vertical axis represents the input port index and the horizontal axis represents the layer index.  Power flows from left to right.  Equalization is achieved, on average, after only around 5 layers given this network with 30 ports.}
\end{figure}

To demonstrate our protocol, we perform numerical simulations of a Clements mesh and optimize it for power equalization from random inputs to uniform outputs. A software package \cite{hughes2018DLA_Control} was written to simulate the mesh and perform the optimizations. This package was written such that it may eventually be augmented to interface with a physical MZI mesh to act as a control mechanism.  The result of 6 independent runs is shown in Fig \ref{fig:equalization}, in which we plot the power in each port within the network after it is optimized using this procedure.

For each run, we initialize a Clements mesh with $N = 30$ input ports and $M = 10$ layers.  Then, we generate a vector of random input powers to couple to the mesh, $P_\textrm{in}$.  When constructing $P_\textrm{in}$, each element is chosen uniformly at random between 0 and 1 and then the whole vector is normalized such that it sums to 1.  The phase of each input mode is set to 0.  For demonstration purposes, we optimize the mesh to output to a uniform output target $T_\textrm{out}$, where each element of $T_\textrm{out}$ is equal to $1/N$, such that both the input and target powers are normalized to each sum to 1.  Although a uniform output target was chosen as it is most applicable to DLA applications, the same protocol may be equally applied to other targets with similar results.

We then step through the mesh from input to output, tuning all MZIs in a given layer according to the protocol introduced in the previous section before moving to the next layer.  To perform the tuning, we use a simple downhill simplex algorithm \cite{avriel_nonlinear_2003} to tune the phase shifters ($\theta$ and $\phi$ in Eq. \ref{eq:U}) in the MZI until they output the correct power as described by the protocol.  From Fig. \ref{fig:equalization}, we notice that equalization is achieved after only around 5 layers on average.

\begin{figure}
\includegraphics[width=1\columnwidth]{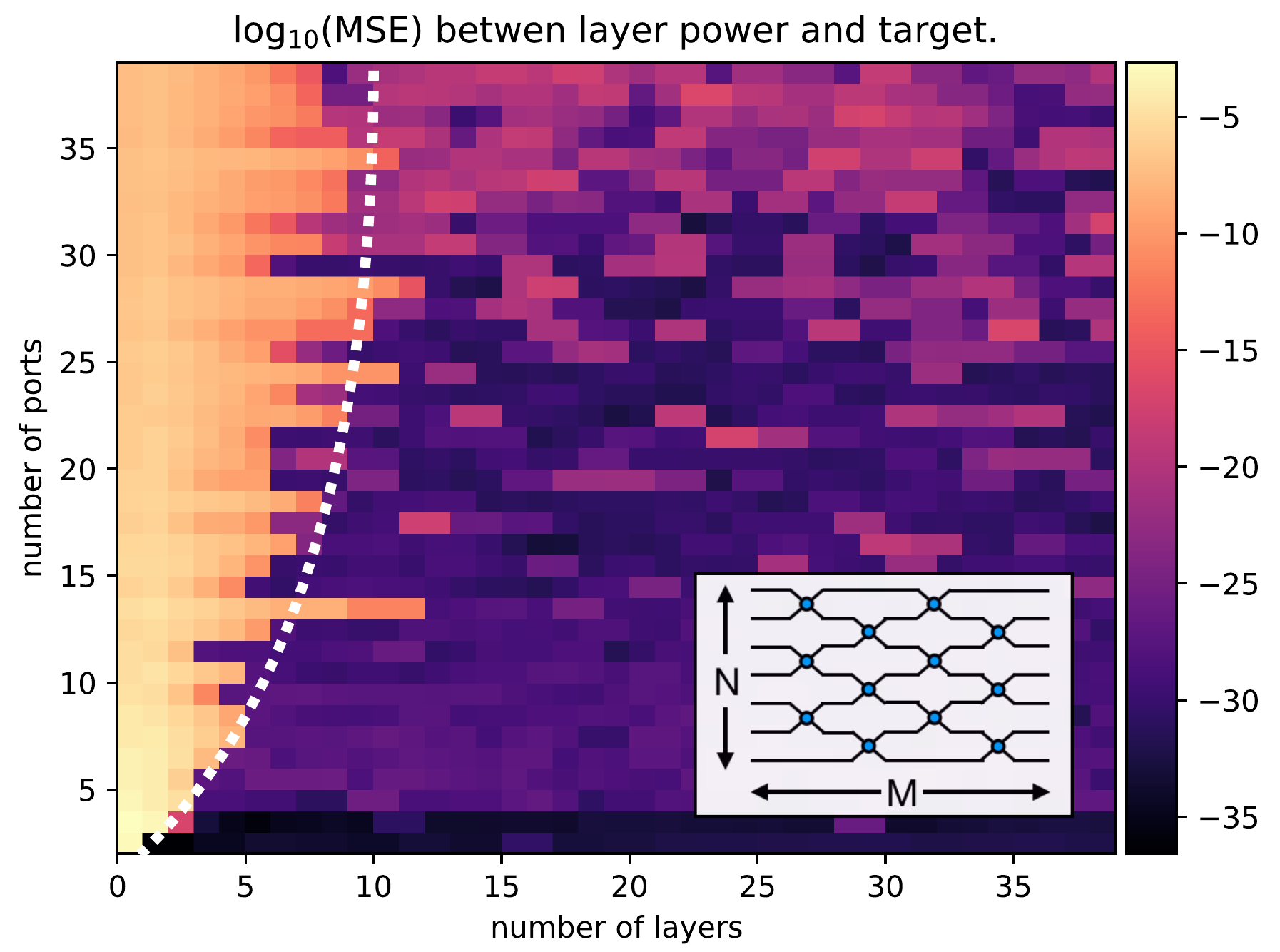}
\caption{\label{fig:scaling} \textbf{Analysis of mesh performance vs. number of input ports}.  We plot the mean-squared-error between the power in each layer and the target power after the mesh has been optimized.  These are averaged over 5 random input values.  We see that the number of layers ($M$) needed to equalize the power in the mesh increases slowly with respect to the number of ports ($N$).}
\end{figure}

To understand how the device operation scales with the number of layers, we studied the performance of the equalization routine as the number of ports are increased.  The results are shown in Fig. \ref{fig:scaling} where we show the mean-squared-error between the power in each layer, defined as 
\begin{equation}
    MSE = \frac{1}{N}\sum_{i=1}^N \left( P_\textrm{in}^{(i)} - T_\textrm{out}^{(i)} \right)^2.
\end{equation}
We then sweep through different mesh sizes and average the data for each mesh over 5 runs with different random inputs.  The number of layers needed for equalization grows slowly as the network size is increased.  This suggests that for very large networks (with $>$ 100-1000 ports), only tens of layers of MZIs may be needed in the Clements mesh to perform power equalization.

\section{\label{sec:DLA_analysis}Application to DLA}
To quantify the benefit of this new power distribution component, we now compare the performance of a DLA with the power splitting approach of Ref. \cite{hughes2018chip} to one with direct coupling and power control from this work and as diagrammed in Fig. \ref{fig:setup}.  For each of these approaches, we give an estimate of the acceleration gradient, number of output waveguides, and acceleration length needed to achieve a given energy gain with a single driving laser.  

Following the analysis of Ref. \cite{hughes2018chip}, we first consider a splitting design in which a single input waveguide is split $N_s$ times to cover the full length of acceleration.  We assume that an optical pulse with energy $\Ud$ and temporal duration $\tau$ is coupled into the input waveguide.  We must ensure that $\Ud$ is below the damage and nonlinearity limit of the waveguides, which was recently measured to be 20 nJ in SiN waveguies using pulses of duration of 250 fs \cite{tan2019silicon}.  We then assume that this initial waveguide is split $N_s$ times, in a binary fashion, to couple to $N = 2^{N_s}$ final waveguides, which are directly coupled to the accelerator. The power coupling efficiency factors corresponding to waveguide splitting and bending loss are denoted by $\eta_s$ and $\eta_b$ respectively. The values used for these efficiencies are given in Table \ref{tab:params} and are consistent with those of Ref. \cite{hughes2018chip}.

Next, we assume each waveguide powers $\Np$ periods of the accelerator, each of height $h$ and width $\beta \lamo$ to satisfy the synchronicity condition, where $\beta$ is the ratio of the electron speed to the speed of light and $\lamo$ is the free space wavelength of the driving laser.  This gives a total acceleration length of
\begin{equation}
    L = N \Np \beta \lamo
    \label{eq:L}
\end{equation}

In terms of these parameters and  the peak electric field of the laser pulse incident on the accelerator channel $(E_0)$, the corresponding power intercepting the DLA is $P_0 \simeq \zeta E_0^2 h N_p \lambda \beta n/ (2 Z_0)$, where $Z_0 = 377 \Omega$ is the impedance of free space, $n$ is the refractive index, and $\zeta$ is a dimensionless factor of order unity (for a Gaussian $\zeta = \pi/2$). Considering losses between the input coupling of a pulse with energy $U_d$ and duration $\tau$, we may compute an expression for $P_0$ and solve for $E_0$ to obtain
\begin{equation}
    \Eo = \sqrt{\frac{2 \Ud Z_0 (\eta_s \eta_b)^{N_s}}{2^{N_s} \zeta \Np \tau \lamo \beta h n}},
    \label{eq:E0}
\end{equation}

The acceleration gradient, defined as the energy gain ($\DE$) per unit length, may be written in terms of the acceleration length $L$ and input electric field $\Eo$ as

\begin{equation}
    G = \frac{\DE}{L} = \kappa \Eo.
    \label{eq:G}
\end{equation}
The proportionality constant or \textit{structure factor} $\kappa$ is a dimensionless quantity that denotes the coupling of incident field to the structure. Structure factors of $\kappa = 0.2$ have been experimentally demonstrated \cite{cesar_optical_2018} and $\kappa = 1.34$ theoretically predicted \cite{barlev:2019} for various structure designs, and by use of a resonant structure with a quality factor $Q$ it can be further enhanced as $\sqrt{Q}$ \cite{hughes2018chip}. We here assume a moderate value of $\kappa = 1$. With the expression above for the gradient, along with Eqs. (\ref{eq:L}) and (\ref{eq:E0}), we may solve for the number of waveguide splits required to accomplish the energy gain of $\DE$ as
\begin{equation}
    N_s = \log_2 \left[ \frac{\DE^2 \tau h n \zeta}{2  \kappa^2 \Ud \Np \beta \lamo Z_0} \right] / \log_2 (2 \eta_s \eta_b)
    \label{eq:N}
\end{equation}
The total number of waveguides feeding the DLA is then $N = 2^{N_s}$. We can see that in the limit where the efficiency terms are equal to unity, $N$ reduces to the quantity contained inside the first logarithm in Eq. (\ref{eq:N}).

In contrast, using a direct coupling method with power distribution components introduced in this work, there is no need for on-chip splitting of the optical power.  Because of this, we may couple $\eta_\textrm{MZI} \times \Ud$ of optical energy separately into each waveguide, where $\eta_\textrm{MZI}$ is the throughput efficiency of the MZI. In the experimental demonstrations of Ref. \cite{annoni_unscrambling_2017,shen_deep_2017}, less than 50\% of power loss was demonstrated for a 4-layer MZI mesh. As a conservative estimate, we assume here $\eta_\textrm{MZI} = 0.25$ for 10 layers.  Following a similar analysis, the electric field incident on the accelerator, given by Eq. (\ref{eq:E0}), loses its $N$ dependence and thus the number of required output waveguides needed scales as 
\begin{equation}
    N^{(\textrm{direct})} =  \frac{\DE}{\kappa} \sqrt{\frac{\zeta \tau h n}{2 \Ud \eta_\textrm{MZI} \Np \beta \lamo Z_0 }}.
    \label{eq:N_direct}
\end{equation}

\begin{figure}[!h]
\includegraphics[width=1\columnwidth]{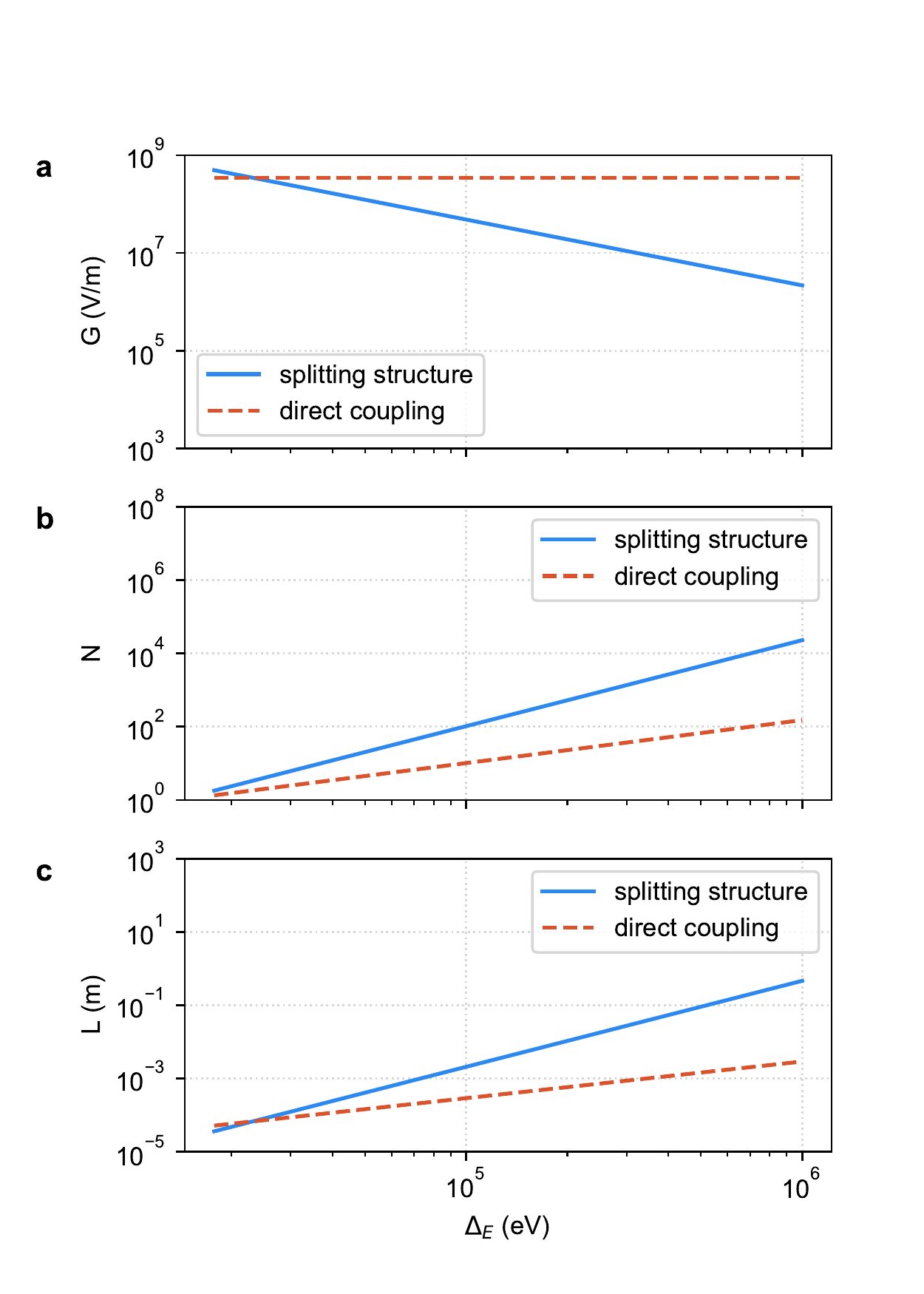}
\caption{\label{fig:plots} \textbf{Figure of merit scaling for different laser coupling architectures.} Solid blue lines refer to the splitting structure of Ref \cite{hughes2018chip} and red dotted lines refer to the structure from this work.  \textbf{a}  The acceleration gradient as a function of electron energy gain.  For the splitting structure, the acceleration gradient diminishes rapidly as the length of acceleration is increased to match the desired energy gain.  However, direct coupling structure in principle achieves uniform gradient.  \textbf{b}  The number of output waveguides ($N$) required to achieve an given energy gain of $\DE$.  \textbf{c}  The acceleration length ($L$) required to achieve an given energy gain of $\DE$.}
\end{figure}

In Fig. \ref{fig:plots} we show the scaling of the acceleration gradient, number of output waveguides, and acceleration length as a function of required energy gain for a DLA given realistic parameters, which are supplied in Table \ref{tab:params}. 

To connect these results with prior work, we note that for the split waveguide approach, Eqs. \ref{eq:L}, \ref{eq:E0}, \ref{eq:G}, \ref{eq:N} give nearly identical results as those of the more involved analysis of \cite{hughes2018chip} for the case of silicon nitride waveguides for a single-stage energy gain of 20 keV, when the same input parameters are used. However, the results in Fig. \ref{fig:plots} for the split waveguide approach (solid blue curves) are somewhat more optimistic than those of the prior work, as recent experiments \cite{tan2019silicon, bar2019design} have provided evidence for larger values of $U_d$ and $\kappa$ than were previously assumed. These more optimistic parameters are reflected in Table \ref{tab:params}. However, due to the $N_s$ scaling of the power loss for the split waveguide approach, higher energy gains per stage rapidly become prohibitive as the resulting gradient drops exponentially and the number of required waveguides increases. In Ref. \cite {hughes2018chip} this difficulty was addressed by cascading multiple laser-coupled stages in series to achieve a reasonable energy gain of 1 MeV.  In contrast, as seen in Fig. \ref{fig:plots} the direct coupling approach can in principle achieve this with one acceleration stage and one input laser, requiring 144 waveguides over a length of 2.8 mm with a gradient of 350 MeV/m. These are promising numbers for a DLA structure and may be improved primarily by optimizing for higher values of $\kappa$ and $\Np$, which may be accomplished using inverse design techniques \cite{hughes_method_2017}.

\begin{table}[!h]
\centering
\begin{tabular}{lccc}
\hline
Metric & Description & Value & Units \\
\hline
$\beta$ & $e^{-}$ speed / $c_0$ & 1 & - \\
$\lamo$ & free space wavelength & 2 & $\mu$m \\
$\Np$ & \# DLA periods / waveguide & 10 & - \\
$\kappa$ & structure coupling factor & 1 & - \\
$\Ud$ & input pulse energy & 20 & nJ \\
$\tau$ & input pulse duration & 250 & fs \\
$h$ & DLA structure height & 2 & $\mu$m \\
$n$ & waveguide refractive index & 2 & - \\
$\eta_s$ & waveguide splitting efficiency & 0.95 & - \\
$\eta_b$ & bend loss efficiency & 0.95 & - \\
$\eta_\textrm{MZI}$ & MZI transmission efficiency & 0.25 & - \\
$\zeta$ & geometrical mode factor & $\pi$/2 & - \\
\hline
\end{tabular}
\caption{\label{tab:params} Set of parameters used in the analysis of Fig. \ref{fig:plots}}.
\end{table}

\section{\label{sec:discussion}Discussion}

Our proposal provides a method for automated control of DLA systems and eliminates the major issues with previous laser coupling schemes.  We show that MZI meshes are a promising candidate for a power distribution system for DLA and our findings may be applied to other applications in integrated optics requiring power routing.  The proposal given here only uses existing optical components and, therefore, should be feasible for an experimental demonstration.

The procedure we introduce for optimizing the MZI mesh to achieve arbitrary power distribution is highly efficient in terms of number of measurements and phase shifter tunings.  For a mesh with $N$ ports and $M$ layers, our protocol replaces one large optimization problem with $2NM$ degrees of freedom to $MN$ independent optimization problems with 2 degrees of freedom each.  This greatly improves the feasibility of implementing this protocol on large meshes with thousands of input ports, for example, which may be needed for large-scale DLA.

We show that a shallow Clements mesh is sufficient for equalizing power from random inputs.  This is of crucial importance for DLA applications where there are limitations on the chip space available for phase shifters and electrical contacts.  It also means that a shorter propagation length within the waveguides can be achieved, which reduces the possible nonlinear effects.  Having a shallow mesh also allows the device to retain a large bandwidth, which decreases with the number of MZIs in an optical path.  This is crucially important for handling sub-picosecond driving pulses used in DLA.  However, the bandwidth of MZI meshes and its scaling with respect to network size is not yet fully understood.  This will need to be tested experimentally or with a separate numerical study.

The optimization protocol presented here decomposes the global optimization problem of tuning the entire mesh into several subproblems involving tuning the individual MZIs.  However, alternatively a gradient-based approach may potentially be used to train the full mesh.  It was shown previously \cite{hughes2018training} that the gradient of the output of an MZI mesh with respect to the dielectric function of each of the phase shifters may be measured experimentally using adjoint fields \cite{veronis_method_2004}. Interestingly, for the case of maximizing the acceleration of a DLA, it was independently shown that the corresponding adjoint fields are given exactly by the fields radiated by the electron beam \cite{hughes_method_2017}.  This suggests an interesting approach to optimizing the MZI mesh towards maximum acceleration by first measuring the radiation from test electron beam, and then using the protocol from Ref. \cite{hughes2018training} to measure the gradient of the acceleration with respect to each of the phase shifters.  With this, one may do parallel, gradient-based updates of the phase shifters and optimize arbitrarily large grids with high efficiency. This idea may be explored in a future study.
\newline

\section{\label{sec:conclusion}Conclusion}

Our study indicates that integrated optical power delivery systems are worth continuing to pursue for DLA.  We presented a path towards automatic power distribution, which is an essential component towards scaling DLA to longer length scales and exciting applications.  We also provide a novel application of the MZI mesh, which is already finding many applications in other exciting reconfigurable optics applications.  Our efficient protocol for optimizing an MZI for arbitrary power distribution may also find many applications beyond DLA.

Integrated optics, and reconfigurable optics in general, allows unique opportunities for accelerators on a chip to take advantage of high precision control and automatic compensation for errors from fabrication, alignment, or drift. This study presents a promising avenue for accomplishing extended acceleration lengths for these accelerators and eventually may enable future applications of DLA technology.

\section{\label{sec:acknowledgements}Acknowledgements}

The authors thank Sunil Pai, Ian Williamson, Momchil Minkov, and Ben Bartlett for useful discussions.

This work is supported by the Gordon and Betty Moore Foundation (GBMF4744).

\bibliography{DLA_Phase_Control}

\end{document}